\documentclass{acm_proc_article-sp}

\usepackage{graphicx}
\usepackage{tabularx}
\usepackage{amsopn}
\usepackage{bm}
\usepackage{booktabs}
\usepackage{epsfig}
\usepackage{subfigure}
\usepackage{amsfonts}
\usepackage{comment}
\usepackage{enumerate}
\usepackage{mathrsfs} 
\usepackage[perpage,symbol]{footmisc}   \setfnsymbol{wiley}   
\usepackage{colortbl} 
\usepackage{caption2} 
\usepackage{etoolbox} 
\patchcmd{\maketitle}{\@copyrightspace}{}{}{} 

\newcommand{\superscript}[1]{\ensuremath{^{\textrm{#1}}}}
\def\sharedaffiliation{\end{tabular}\newline\begin{tabular}{c}}
\def\wu{\superscript{1}}
\def\wg{\superscript{2}}
\def\ws{\superscript{3}}

\begin{document}
%
\conferenceinfo{WWW}{'13 Rio de Janeiro, Brazil}

\title{Extending modularity by capturing the similarity attraction feature in the null model}
\numberofauthors{3}

\author{
  \alignauthor \ Xin Liu \wu\superscript{,}\wg\superscript{,}\ws\\
  \email{tsinllew@ai.cs.titech.ac.jp}
  \alignauthor \ Tsuyoshi Murata \wu\\
  \email{murata@cs.titech.ac.jp}
  \alignauthor \ Ken Wakita \wu\titlenote{To whom correspondence should be addressed.}\\
  \email{wakita@is.titech.ac.jp}
  \sharedaffiliation
  \begin{tabular}{ccccc}
    \affaddr{{\wu}Tokyo Institute of Technology{\ }} & & \affaddr{{\wg}CREST, JST{\ }} & & \affaddr{{\ws}Wuhan University of Technology{\ }} \\
    \affaddr{2-12-1 Ookayama}                        & & \affaddr{K's Gobancho, 7, Gobancho}           & & \affaddr{122 Luoshi Road} \\
    \affaddr{Meguro, Tokyo}                          & & \affaddr{Chiyoda, Tokyo}                      & & \affaddr{Wuhan, Hubei} \\
    \affaddr{152-8552 Japan}                         & & \affaddr{102-0076 Japan}                      & & \affaddr{430070 China} \\
  \end{tabular}
}
\maketitle
\begin{abstract}

Modularity is a widely used measure for evaluating community structure in networks. The definition of modularity involves a comparison of within-community edges in the observed network and that number in an equivalent randomized network. This equivalent randomized network is called the null model, which serves as a reference. To make the comparison significant, the null model should characterize some features of the observed network. However, the null model in the original definition of modularity is unrealistically mixed, in the sense that any node can be linked to any other node without preference and only connectivity matters. Thus, it fails to be a good representation of real-world networks. A common feature of many real-world networks is ``similarity attraction'', i.e., edges tend to link to nodes that are similar to each other. We propose a null model that captures the similarity attraction feature. This null model enables us to create a framework for defining a family of Dist-Modularity adapted to various networks, including networks with additional information on nodes. We demonstrate that Dist-Modularity is useful in identifying communities at different scales.
\end{abstract}

\section{Introduction}
Many social, biological, and information systems can be well described by networks, where nodes represent fundamental entities of the system, such as individuals, users, genes, web pages, and so on, and edges represent relations or interactions between the entities \cite{NewmanNetworksAnIntroduction}. In recent years, there has been a surge of interest in the analysis of networks. A highly discussed topic is community detection --- The identification of groups of network nodes, known as communities, within which edges are dense, but between which edges are sparse \cite{GirvanNewmanCommuityDefinition}. Community detection is considered as a crucial step towards inferring function units of the underlying system, such as collections of pages on closely related topics on the web \cite{FlakeWebCommunities} or groups of people with common interests in social media \cite{NewmanModularityDefinition}. Thus it provides insight into how the system is internally organized and works.

In practice, optimization methods are widely used for community detection \cite{FortunatoCommunityDetectionReview}. The basic idea is to define a quantity measure for evaluating the ``goodness'' of a partition of a network into communities, and then to search through possible partitions for the one with the highest score. A variety of partition measures have been proposed, but the most famous one is known as the modularity \cite{NewmanModularityDefinition}.

The spirit behind modularity is that a good partition of a network into communities is the one in which the number of within-community edges is larger than that number in some equivalent randomized network. This equivalent randomized network is called the null model, which serves as a reference. To make the comparison significant, the null model should characterize some features of the observed network. The fundamental requirement is that the number of nodes and the expected number of edges are the same as the observed network.

Beyond the fundamental requirement, there are many possible choices of the null model. A simple but less used choice is the uniform model, where edges are placed with equal probability between all node pairs. In addition, a popular choice proposed by Newman and Girvan \cite{NewmanModularityDefinition} is the configuration model \cite{NewmanRandomGraphArbitraryDegree, ChungRandomGraphsExpectedDegree}, where the expected degree of each node is equal to the degree of the corresponding node in the observed network.

However, both the uniform model and the configuration model are not good representations of many real-world networks. The uniform model has a binomial degree distribution, which is entirely unlike the right-skewed degree distributions found in most real-world networks \cite{BarabasiScalingInRandomNetwork, AmaralSmallWorldNetworks}. The configuration model is unrealistically mixed, in the sense that any node can be linked to any other node without preference and only connectivity matters. Consequently, both models may fail to be a valuable reference and result in less accurate modularity \footnote{An evidence is that although it is axiomatically recognized that a Erd{\H o}s-R{\'e}nyi random graph \cite{ERModel} has no community structure, people surprisingly found that it has partitions with high modularity, which is based on adopting configuration model as the null model \cite{GuimeraRandomGraphModularity, ReichardtTrulyModular}.}.

A common feature of many real-world networks is ``similarity attraction'', i.e., edges tend to link to nodes that are similar to each other. Two nodes are considered to be similar if they have much common information. For instance, in social networks, two individuals with many common friends are more likely to be friends themselves; individuals with similar interests are more likely to be friends. On the one hand, a simple and effective way to measure such similarity is based solely on the network structure, known as structural similarity \cite{TaoEpidemicDynamics}. On the other hand, additional information on nodes, such as attributes and contents can also be used to measure such similarity.

In this paper, we propose a new null model that captures the similarity attraction feature. Taking this null model as a reference to compare with the observed network, we create a framework for generating a family of modularity that are adapted to various networks. Since we employ appropriate distance functions to estimate the node similarity, we call this family of modularity as Dist-Modularity. Within this framework, we have the following interesting findings
\begin{itemize}
\item Dist-Modularity incorporates NG-Modularity (Newman and Girvan's modularity that takes the configuration model as the null model) as a special case;
\item Dist-Modularity is suitable for networks with additional information on nodes, with attributes or contents being active in the estimation of node similarity;
\item Optimizing Dist-Modularity brings community structure at different scales;
\item Dist-Modularity provides an in-depth view of the close relationship between graph partitioning/spectral clustering and community detection.
\end{itemize}

The rest of the paper is organized as follows: Section~\ref{sec2} gives a survey of modularity. Section~\ref{sec3} introduces our null model as well as Dist-Modularity. Section~\ref{sec4} presents experimental results. Section~\ref{sec5} reviews related research, followed by a conclusion in Section~\ref{sec6}.

\section{Modularity}\label{sec2}
In this section, we give a survey of modularity, which is the basis of the following discussion. As mentioned above, modularity involves a comparison between the observed network and a reference network, the null model. Formally, modularity is defined to be the fraction of within-community edges in the observed network minus the expected value of that fraction in the null model. In a mathematical expression, modularity reads
\begin{eqnarray}
\text{Q}(\bm{\mathscr{C}})=\frac{1}{2m} \sum_{i,j=1}^n (A_{ij}-P_{ij})\,\delta(l_{i},l_{j}), \label{eq1}
\end{eqnarray}
where $\bm{\mathscr{C}}$ is a partition represented as a community assignment vector on the right-hand side of the equation, with element $l_{i}$ indicating the community membership of the $i$th node $v_i$, $n$ is the number of nodes, $m$ is the number of edges, ${A}_{ij}$ is the element of the adjacency matrix $\mathbf{A}$ representing the number of edges between $v_i$ and $v_j$ in the observed network, ${P}_{ij}$ is the expected value of that number in the null model, and $\delta$ is the Kronecker's delta.

To make Eq.\,\eqref{eq1} significant, we require that the expected number of edges in the null model equal the number of edges in the observed network. That is,
\begin{eqnarray}
\sum_{i,j=1}^n P_{ij} = \sum_{i,j=1}^n A_{ij} = 2m. \label{eq2}
\end{eqnarray}
Then, it is clear that $\text{Q} \in [-1,1]$.

Note that the so called modularity matrix $\mathbf{B}$ \cite{NewmanModularityFurtherAnalysis, NewmanModularityEigenvectors} is defined as
\begin{eqnarray}
\mathbf{B} = \mathbf{A}-\mathbf{P}. \label{eq3}
\end{eqnarray}
According to Eq.\,\eqref{eq2}, all elements of the modularity matrix sum to zero, which means that modularity of the partition that places all nodes in a single community is always zero.

Apart from the constraint in Eq.\,\eqref{eq2}, there is some freedom about choosing the null model. Matching the null model with the uniform model, where edges are placed with equal probability between all node pairs, we have
\begin{eqnarray}
P_{ij}^{\text{Unif}} = 2m/n^2. \label{eq4}
\end{eqnarray}
Correspondingly, we obtain the Unif-Modularity
\begin{eqnarray}
\text{Q}^{\text{Unif}}(\bm{\mathscr{C}})=\frac{1}{2m} \sum_{i,j=1}^n (A_{ij}-P_{ij}^{\text{Unif}})\,\delta(l_{i},l_{j}). \label{eq5}
\end{eqnarray}

Matching the null model with the configuration model, which preserves the degree sequence of the observed network and place edges randomly, we have \cite{NewmanModularityFurtherAnalysis, NewmanModularityEigenvectors}
\begin{eqnarray}
P_{ij}^{\text{NG}}=k_ik_j/2m, \label{eq6}
\end{eqnarray}
where $k_i=\sum_{j=1}^n A_{ij}$ is the degree of $v_i$.
Correspondingly, we obtain the famous NG-Modularity
\begin{eqnarray}
\text{Q}^{\text{NG}}(\bm{\mathscr{C}})=\frac{1}{2m} \sum_{i,j=1}^n (A_{ij}-P_{ij}^{\text{NG}})\,\delta(l_{i},l_{j}), \label{eq7}
\end{eqnarray}
which is proposed by Newman and Girvan and is widely used in practice.

\section{Dist-Modularity}\label{sec3}
As pointed out previously, a common feature of many real-world networks is similarity attraction. However, neither the uniform model nor the configuration model can capture this feature. In this section, we propose a new null model with this feature built in. This new null model enables us to define a family of Dist-Modularity adapted to various networks.

We use $d_{ij}$ to denote the similarity distance between $v_i$ and $v_j$ --- The smaller of $d_{ij}$, the more similar of the two nodes. $d_{ij}$ can be estimated by a distance function that takes some available information about $v_i$ and $v_j$ as input, e.g., the network structure information (such as the neighbors of $v_i$ and $v_j$), additional information (such as attributes and contents of $v_i$ and $v_j$), or a mixture of them. Generally, $d_{ij}$ should satisfy the following constraints
\begin{eqnarray}
d_{ij} \geq 0 \text{ with equality iff } i=j, \label{eq8}
\end{eqnarray}
\begin{eqnarray}
d_{ij} = d_{ji}, \label{eq9}
\end{eqnarray}
\begin{eqnarray}
d_{ij} \leq d_{it} + d_{tj}. \label{eq10}
\end{eqnarray}
Eq.\,\eqref{eq8} is self-evident --- A node is the most similar to itself and hence the distance is zero. Eqs.\,\eqref{eq9} and \eqref{eq10} indicate the triangle inequality and the symmetry constraints, respectively. We note that estimating $d_{ij}$ is not the primary focus of this paper. Many related research have been conducted. For interested readers, please refer to Ref. \cite{LvRecommenderSystems, ZhouLinkPredictionReview} for more details.

In the following, we first develop the new null model, then we show how it can be generalized to produce a family of Dist-Modularity. For simplicity and convenience, we limit our vision to undirected networks. Nonetheless, our developments and conclusions can be easily extended to directed networks, as done for NG-Modularity \cite{LeichtCommunityDirectedNetwork}.

\subsection{The New Null Model}\label{sec3.1}
In our null model, the edges are placed at random and the expected number of edges between $v_i$ and $v_j$ is
\begin{eqnarray}
P_{ij}^{\text{Dist}} = (\tilde{P}_{ij} + \tilde{P}_{ji}) / 2, \label{eq11}
\end{eqnarray}
\begin{eqnarray}
\tilde{P}_{ij} = \frac { k_ik_j\,e^{-(d_{ij}/\sigma)^2} } { \sum_{t=1}^n k_t\,e^{-(d_{ti}/\sigma)^2} }, \label{eq12}
\end{eqnarray}
where $\sigma \in (0,+\infty)$ is a parameter which will be explained later.

\subsubsection{Interpretation}\label{sec3.1.1}
\begin{figure}[!t]
\centering
\includegraphics[width=0.50\textwidth]{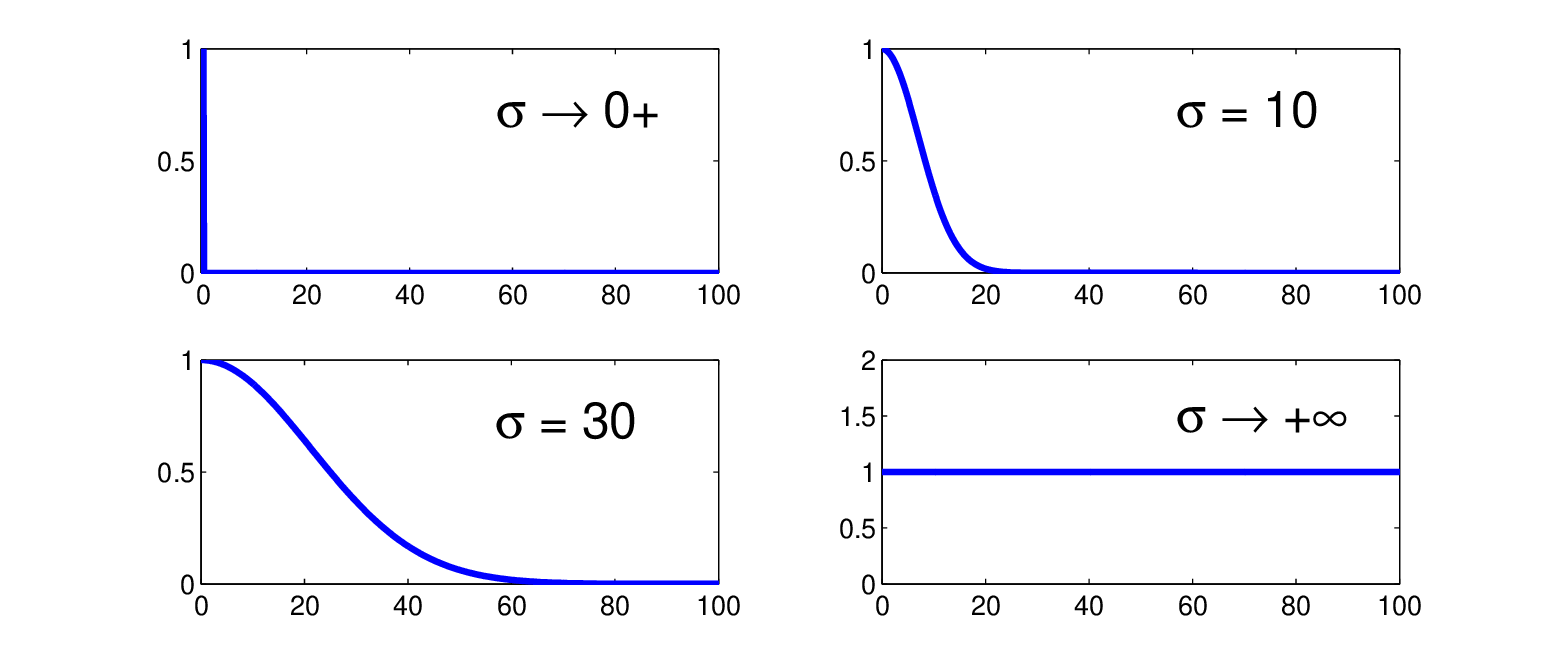}
\caption{\label{fig1} The plot of function $f(d)=e^{-(d/\sigma)^2}$ with different values of $\sigma$.}
\end{figure}

Eq.\,\eqref{eq12} can be interpreted by the data field idea proposed by Li \cite{LiAIUncertainty}, which introduces the field theory in Physics for describing interactions between particles \footnote{From the classic concept of field introduced by M. Faraday in 1837, the field as an interpretation of non-contact interaction between particles at different granularity levels, from atom to universe, has achieved great success.} into the data space. Let us suppose each node exert forces on others by generating a field. The conservative field theory says that the potential at a point in space is directly proportional to the ``power'' of the field source \footnote{We use the term ``power'' to unify a concept that has different meanings in different fields. For instance, in the gravitational field it means the mass of a particle; in the electrostatic field it means the charge of a particle.}, and decreases as the distance to the source increases. Based on this, we can calculate the potential at $v_j$ of the field driven by $v_i$ as
\begin{eqnarray}
\varphi_{i}(j)=k_i\,e^{-(d_{ij}/\sigma)^2}. \label{eq13}
\end{eqnarray}
where the power of $v_i$, denoted by $N_i$, is supposed to be the node degree $k_i$, and the deterrence function $f(d)=e^{-(d/\sigma)^2}$, falling in $(0,1]$, monotonically decreases with $d$. The parameter $\sigma \in (0,+\infty)$ reflects the interaction range of the field --- If $\sigma$ is small, $f(d)$ decreases sharply, indicating a short-range field; If $\sigma$ is large, $f(d)$ decreases slowly, indicating a long-range field (see Fig.~\ref{fig1}). Note that the fields driven by different nodes can be superposed. Suppose these fields are scalar fields, then the superposed potential at $v_j$ is
\begin{eqnarray}
\varphi_{\text{sup}}(j) = \sum_{t=1}^n \varphi_{t}(j) = \sum_{t=1}^n k_t\,e^{-(d_{tj}/\sigma)^2}. \label{eq14}
\end{eqnarray}
Then Eq.\,\eqref{eq12} can be rewritten as
\begin{eqnarray}
\tilde{P}_{ij} = \frac { \varphi_{i}(i)\,\varphi_{j}(i) } { \varphi_{\text{sup}}(i) }, \label{eq15}
\end{eqnarray}
where we have used $d_{ii}=0$ and $d_{ij}=d_{ji}$. That is, $\tilde{P}_{ij}$ is the product of potentials at $v_i$ of the fields driven by $v_i$ and $v_j$ separately, divided by the superposed potential at $v_i$.

\subsubsection{Properties}\label{sec3.1.2}
In the following, we describe the properties of our null model. First, it can be found that
\begin{eqnarray}
P_{ij}^{\text{Dist}} = P_{ji}^{\text{Dist}}. \label{eq16}
\end{eqnarray}
Eq.\,\eqref{eq16} implies that edges in our null model are undirected, a constraint for representing undirected networks. Second, from Eq.\,\eqref{eq12} it is easy to derive that
\begin{eqnarray}
\sum_{i,j=1}^n \tilde{P}_{ij} = \sum_{i,j=1}^n \tilde{P}_{ji} = \sum_{i=1}^n k_i = 2m, \label{eq17}
\end{eqnarray}
and hence
\begin{eqnarray}
\sum_{i,j=1}^n P_{ij}^{\text{Dist}} = \sum_{i,j=1}^n A_{ij} = 2m. \label{eq18}
\end{eqnarray}
That is, the number of edges of the observed network is preserved, a basic requirement for any null model as given in Eq.\,\eqref{eq2}. Third, Eq.\,\eqref{eq12} tells us that $\tilde{P}_{ij}$ is positively related to $k_i$ and negatively related to $d_{ij}$. In other words, edges tend to link to high degree nodes and nodes that are similar to each other. Therefore, our null model captures similarity attraction feature of many real-world networks, while respecting the degree factor \footnote{In fact, our null model has a degree distribution that is approximately the same as that of the observed network.}.

\begin{figure*}[!t]
\centering
\includegraphics[width=0.95\textwidth]{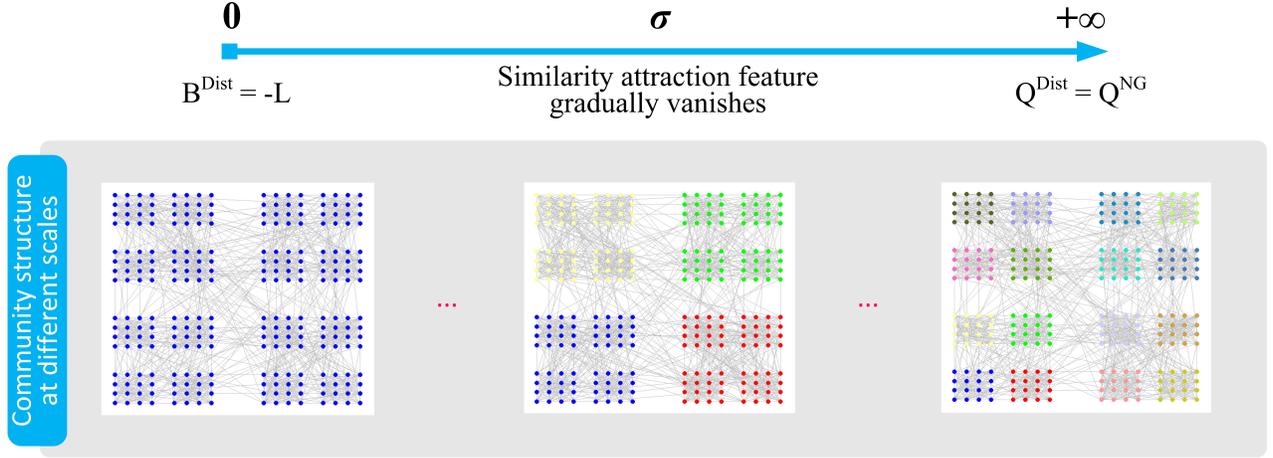}
\caption{\label{fig2} The outcomes related to Dist-Modularity as $\sigma$ increases from $0$ to $+\infty$.}
\end{figure*}

\subsection{Dist-Modularity}\label{sec3.2}
Based on our null model, we can define Dist-Modularity as
\begin{eqnarray}
\text{Q}^{\text{Dist}}(\bm{\mathscr{C}},\sigma)=\frac{1}{2m} \sum_{i,j=1}^n (A_{ij}-P_{ij}^{\text{Dist}})\,\delta(l_{i},l_{j}). \label{eq19}
\end{eqnarray}
Note that there is a range parameter $\sigma\in(0,+\infty)$ in $P_{ij}^{\text{Dist}}$, and hence different values of $\sigma$ bring different Dist-Modularity.

In one extreme case, $\sigma \to 0+$, the range of the field driven by each node is so short that the potential only exists at the source node. This gives
\begin{equation}
\lim\limits_{\sigma \to 0+} P_{ij}^{\text{Dist}} =
\begin{cases}
k_i,   & \text{if}\ i=j; \\
0,     & \text{otherwise}.
\end{cases}
\label{eq20}
\end{equation}
As a result, each node only links to its most similar node, i.e., itself. It is interesting to notice that
\begin{eqnarray}
\lim\limits_{\sigma \to 0+} \left( A_{ij} - P_{ij}^{\text{Dist}} \right) = A_{ij}-k_i\,\delta(i,j). \label{eq21}
\end{eqnarray}
Eq.\,\eqref{eq21} implies that the Laplacian matrix $\mathbf{L}$, the key matrix in graph partitioning \cite{KernighanGraphPartitioning} and spectral clustering \cite{LuxburgSpectralClusteringTutorial}, and the modularity matrix $\mathbf{B}$ are unified as $\sigma$ approaches $0$ --- The only difference is the sign.

In the other extreme case, $\sigma \to +\infty$, the range of the field driven by each node is so long that the potential at each node equals. This gives
\begin{eqnarray}
\lim\limits_{\sigma \to +\infty} P_{ij}^{\text{Dist}} = k_ik_j/2m. \label{eq22}
\end{eqnarray}
As a result, there is no similarity attraction effect and Dist-Modularity reduces to NG-Modularity.

As $\sigma$ increases from $0$ to $+\infty$, the range of the field becomes longer and longer. Correspondingly, Dist-Modularity indicates the scenario where the similarity attraction effect in the null model gradually vanishes.

It is worthwhile to notice that, as $\sigma$ increases from $0$ to $+\infty$, optimizing dist-modularity brings community structure at different scales. First, $\lim\limits_{\sigma \to 0+} \text{Q}^{\text{Dist}}$ can be rewritten as
\begin{eqnarray}
\lim\limits_{\sigma \to 0+} \text{Q}^{\text{Dist}}(\bm{\mathscr{C}},\sigma)
= \frac{1}{2m} \sum_{g=1}^c \sum_{i,j \in C_g} A_{ij} - 1 , \label{eq23}
\end{eqnarray}
where $c$ is the number of communities, $C_g$ is the $g$th community. Eq.\,\eqref{eq23} implies that $\lim\limits_{\sigma \to 0+} \text{Q}^{\text{Dist}}$ is optimized to $0$ when the network is divided into a single community or several communities corresponding to its connected components. Obviously this is the community structure at the coarse scale. Second, optimizing $\lim\limits_{\sigma \to +\infty} \text{Q}^{\text{Dist}}$ results in community structure whose scale is the same as that of optimizing NG-Modularity. Third, as $\sigma$ ranges from $0$ to $+\infty$, optimizing $\text{Q}^{\text{Dist}}$ brings multi-scale community structures. All of the outcomes in this section are summarized in Fig.~\ref{fig2}.

\subsection{Dist-Modularity Family}\label{sec3.3}
So far we have proposed a new null model that captures the similarity attraction feature. The corresponding Dist-Modularity only applies to plain networks, i.e., networks with only topology information. Many networks in real world have additional information on nodes, such as the geographical position of a location, profile of a person, or contents of a web page. In the following, we show how this null model can be generalized to produce a family of Dist-Modularity for various networks.

Now let us look back at Eqs.\,\eqref{eq13} and \eqref{eq14}, the expression of the potentials, which is the core component of our null model. It can be found that the power of the field source $N_i$ and the deterrence function $f(d)$ are specifically specified as the node degree $k_i$ and $e^{-(d/\sigma)^2}$, respectively. However, we have many more choices while preserving the desired properties of the null model. Note that the general expressions of Eqs.\,\eqref{eq13} and \eqref{eq14} are
\begin{eqnarray}
\varphi_{i}(j)=N_i\,f(d_{ij}), \label{eq24}
\end{eqnarray}
\begin{eqnarray}
\varphi_{\text{sup}}(j) = \sum_{t=1}^n N_t\,f(d_{tj}). \label{eq25}
\end{eqnarray}
Substituting Eqs.\,\eqref{eq24} and \eqref{eq25} into Eq.\,\eqref{eq15}, we have
\begin{eqnarray}
\tilde{P}_{ij} = \frac { N_iN_jf(d_{ij})} { \sum_{t=1}^n N_tf(d_{ti}) }. \label{eq26}
\end{eqnarray}
Finally, with Eqs.\,\eqref{eq11} and \eqref{eq26} we can derive that
\begin{eqnarray}
\sum_{i,j=1}^n P_{ij}^{\text{Dist}} = \sum_{i=1}^n N_i. \label{eq27}
\end{eqnarray}
As indicated in Eq.\,\eqref{eq2}, a fundamental requirement for a null model is that the number of edges equal that number in the observed network. Eq.\,\eqref{eq27} implies that this requirement can be satisfied if $\sum_{i=1}^n N_i=2m$. This can be easily achieved by normalizing $N_i$ as
\begin{eqnarray}
N_i = \frac{N_i}{\sum_{i=1}^n N_i}2m. \label{eq28}
\end{eqnarray}

Due to the above developments, there is a large freedom in specifying $N_i$ and $f(d)$. For example, $N_i$ can be equally distributed among all nodes, i.e., specifying $N_i=2m/n$. In networks with additional information on nodes, $N_i$ can be specified as the most representative attribute of $v_i$, or a comprehensive index obtained from several attributes of $v_i$. As for $f(d)$, we list several possible choices as given in Ref. \cite{LiAIUncertainty}
\begin{eqnarray}
f(d)=1/\big(1+(d/\sigma)^2\big), \label{eq30}
\end{eqnarray}
\begin{eqnarray}
f(d)=1, \label{eq31}
\end{eqnarray}
\begin{equation}
f(d) =
\begin{cases}
1,   & \text{if}\ d\leq\sigma; \\
0,     & \text{otherwise},
\end{cases}
\label{eq32}
\end{equation}
where $\sigma \in (0,+\infty)$ are parameters. Eq.\,\eqref{eq30} indicates a gravitational-like field ($\sigma$ reflects the interaction range of the field) and the corresponding null model is similar to the one developed in Section~\ref{sec3.1}. Eq.\,\eqref{eq31} indicates a flat field so that the similarity attraction feature vanishes. Eq.\,\eqref{eq32} indicates a square-well field so that any node in the corresponding null model can only link to nodes at a distance smaller than $\sigma$.

Eqs.\,\eqref{eq30}-\eqref{eq32} indicate monotonically decreasing functions. However, this is not a necessary constraint. $f(d)$ can be a function that monotonically increases with $d$. This results in a null model in which edges between dissimilar nodes are facilitated and edges between similar nodes are restricted. Besides, $f(d)$ can be learned from the observed network as
\begin{eqnarray}
f(d)=\big( \sum_{
  \substack{i,j=1\\
            d_{ij}=d}
  }^n A_{ij} \big) / 2m, \label{eq33}
\end{eqnarray}
with a binning procedure for smoothness \cite{ExpertSpatialCommunity}.

The freedom in specifying $N_i$ and $f(d)$ enables us to create a framework to produce a family of Dist-Modularity adapted to various networks. Within this framework, it is interesting to see some relations with previous work. As listed below, we can exactly recover Unif-Modularity and NG-Modularity as special cases of Dist-Modularity, with appropriate specifications of $N_i$ and $f(d)$.
\begin{itemize}
\item Specifying $N_i=2m/n$ and $f(d)=1$, we have $P_{ij}^{\text{Dist}} = 2m/n^2$ and thus $\text{Q}^{\text{Dist}}=\text{Q}^{\text{Unif}}$;
\item Specifying $N_i=k_i$ and $f(d)=1$, we have $P_{ij}^{\text{Dist}} = k_ik_j/2m$ and thus $\text{Q}^{\text{Dist}}=\text{Q}^{\text{NG}}$.
\end{itemize}

\begin{figure*}[!t]
\centering
\subfigure[][]{\label{fig3a}
\includegraphics[width=0.45\textwidth]{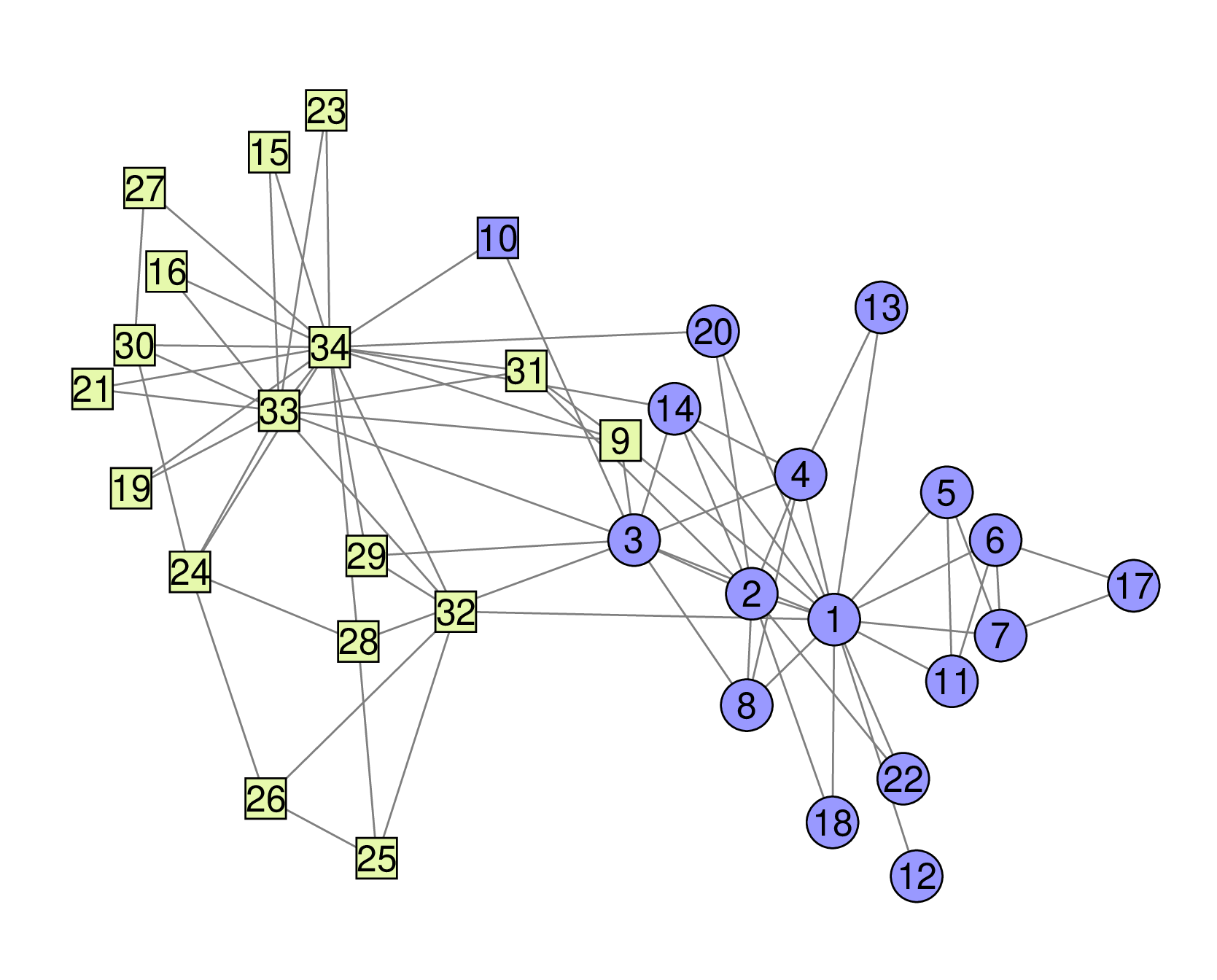}
}
\subfigure[][]{\label{fig3b}
\includegraphics[width=0.45\textwidth]{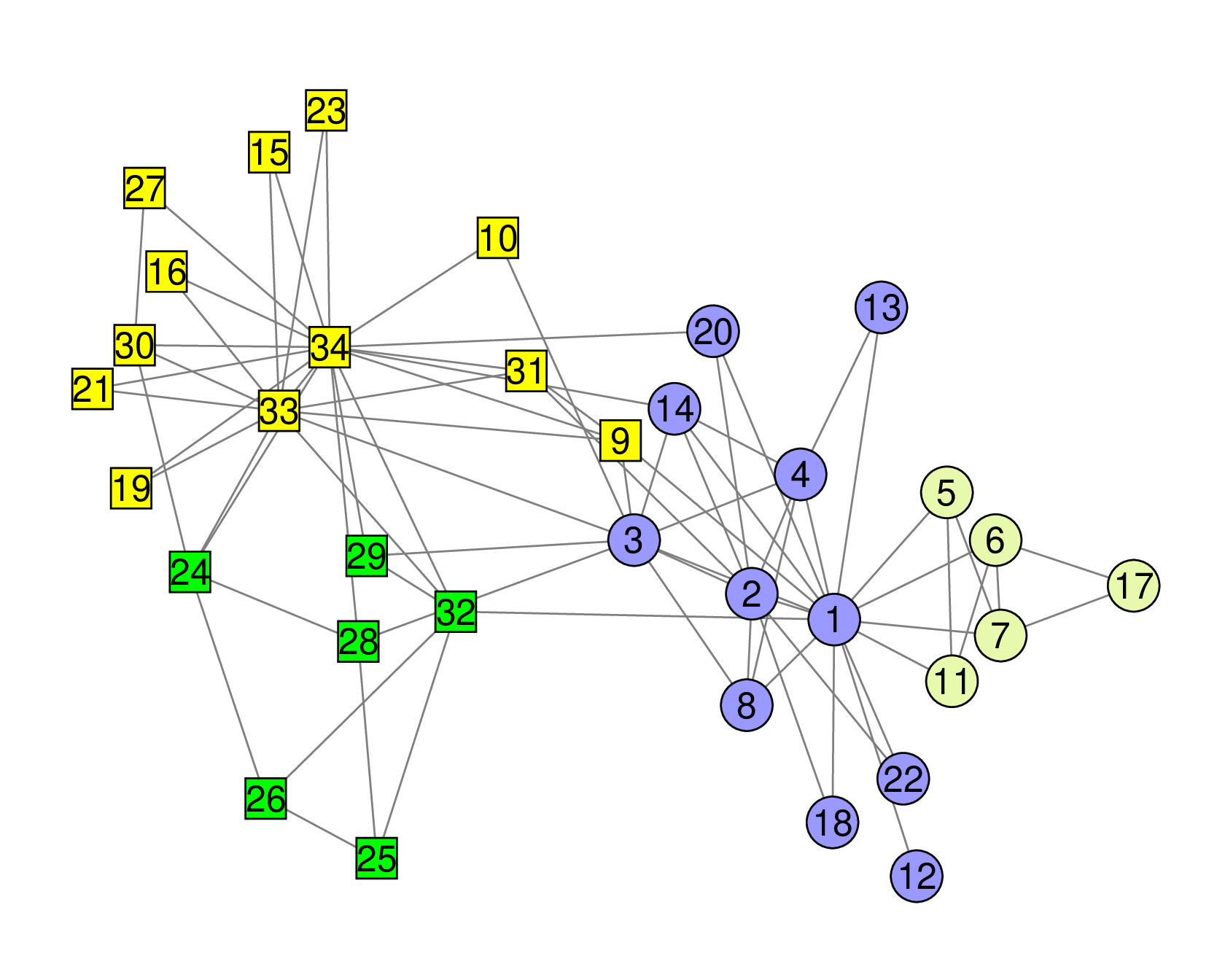}
}
\caption{\label{fig3} Community structure of the Zachary's karate club network obtained by Dist-Modularity optimization algorithm, with $N_i=k_i$ and $f(d)=e^{-(d/\sigma)^2}$. Nodes in the same community are painted in the same color. The true partition of the network into two groups is represented as square and circle symbols. (a) Community structure obtained when $\sigma \in [0.61\bar{d}, 0.64\bar{d}]$. (b) Community structure obtained when $\sigma \in [2\bar{d}, +\infty]$. $\bar{d}$ is the average distance between all node pairs.}
\end{figure*}

\section{Results}\label{sec4}
In this section, we show community detection results by optimizing Dist-Modularity in three representative social networks: the Zachary's karate club network \cite{ZacharyNetwork}, the bottlenose dolphin network \cite{LusseauDolphinNetwork1, LusseauDolphinNetwork2}, and the antenna-to-antenna network of D4D mobile phone datasets \cite{D4DData}.

\subsection{Optimization Algorithm}\label{sec4.1}
Any algorithm designed for optimizing NG-Modularity can be modified to optimize Dist-Modularity. Generally, finding the global optimal solution is NP-complete \cite{BrandesModularityNPCompleteness}, and different heuristics may search over different solution subspaces. Here we consider three popular algorithms: 1) the greedy agglomerative algorithm \cite{NewmanGreedy, ClausetFastGreedy}, 2) the modularity-specialized label propagation algorithm \cite{LiuLPAmplus}, and 3) the Louvain algorithm \cite{BlondelFastUnfolding} plus node moving refinement steps. We take the best solution from those obtained by the three algorithms.

\subsection{Zachary's Karate Club Network}\label{sec4.2}
First, we look at the Zachary's karate club network. This network accounts for a three-year study of the friendships between 34 members of a karate club at a university in 1970. At some point during the study, a conflict between the club administrator and the instructor led to the fission of the club into two separate groups, supporting the administrator and the instructor, respectively. As shown in Fig.~\ref{fig3a}, the administrator and the instructor are denoted as nodes 34 and 1, and the two groups are represented as square and circle symbols, respectively. This network is regularly used as a benchmark to test community detection algorithms, and the challenge is whether an algorithm can obtain the true partition of the two groups from the original network structure.

\begin{figure*}[!t]
\centering
\subfigure[][]{\label{fig4a}
\includegraphics[width=0.48\textwidth]{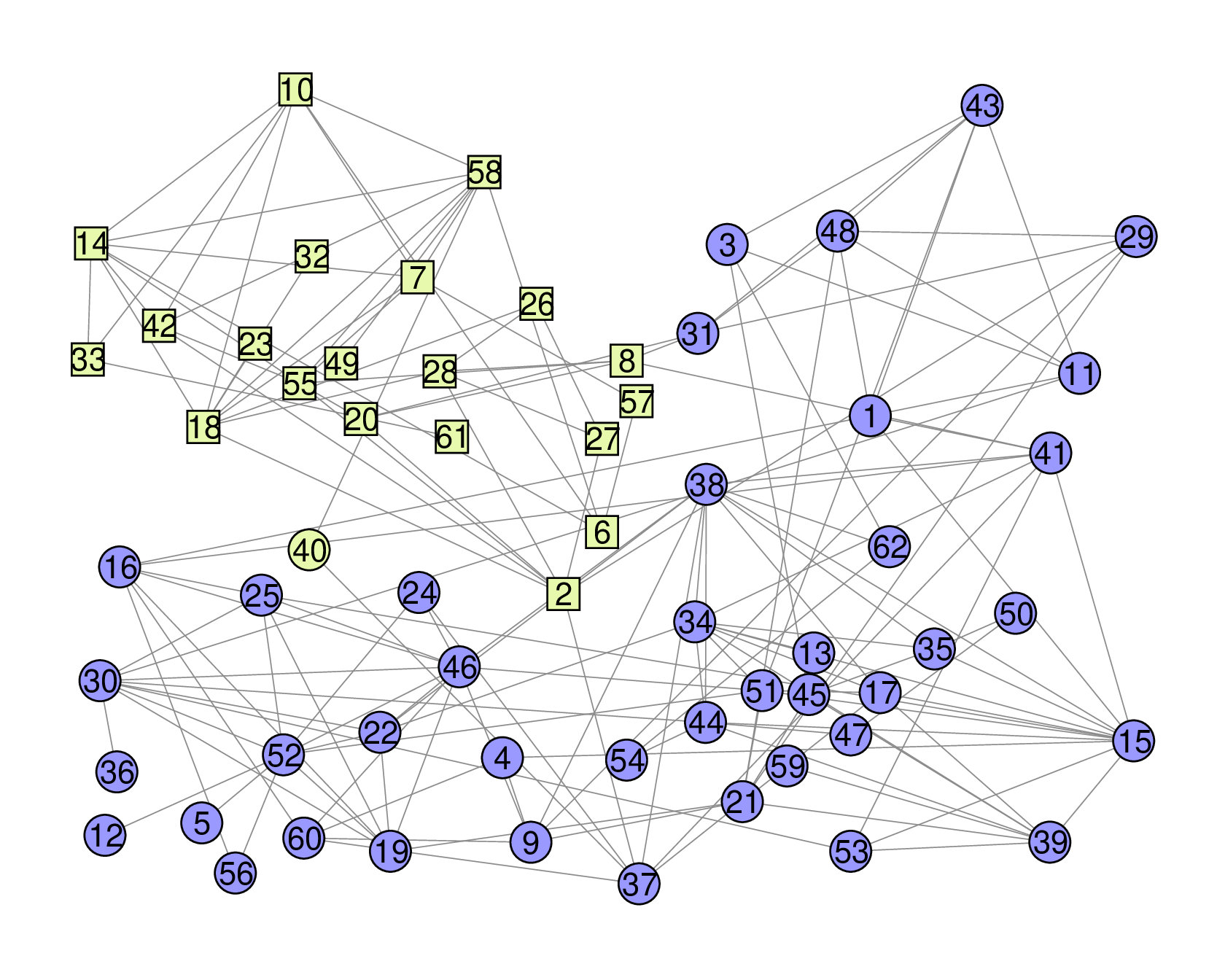}
}
\subfigure[][]{\label{fig4b}
\includegraphics[width=0.48\textwidth]{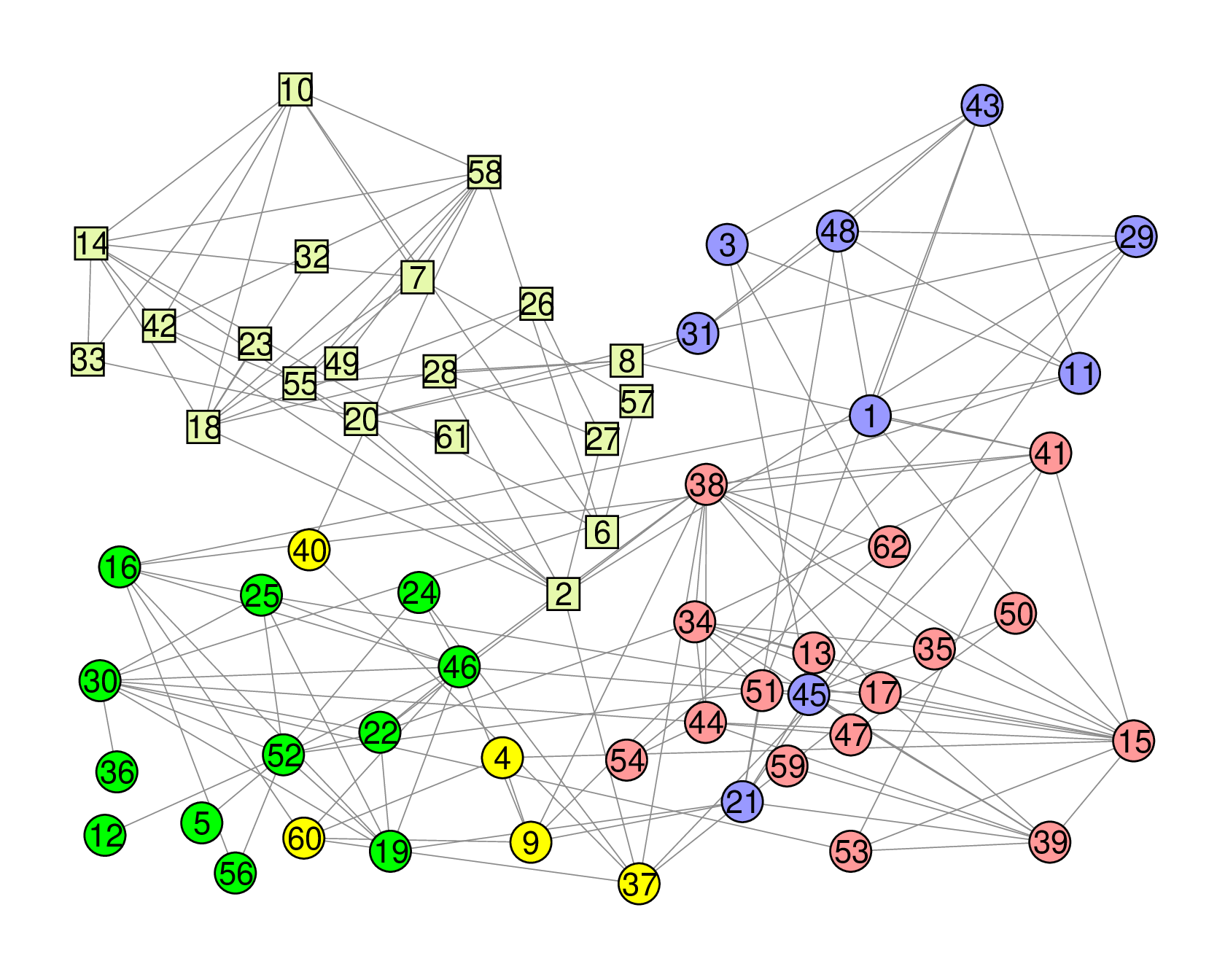}
}
\caption{\label{fig4} Community structure of the bottlenose dolphin network obtained by Dist-Modularity optimization algorithm, with $N_i=k_i$ and $f(d)=e^{-(d/\sigma)^2}$. Nodes in the same community are painted in the same color. The true partition of the network into two groups is represented as square and circle symbols. (a) Community structure obtained when $\sigma=0.5\bar{d}$. (b) Community structure obtained when $\sigma \in [0.9\bar{d}, +\infty]$. $\bar{d}$ is the average distance between all node pairs.}
\end{figure*}

The best partition of this network in terms of optimizing NG-Modularity is the one into four communities as shown in Fig.~\ref{fig3b} (Nodes in the same community are painted in the same color.), with the value of $\text{Q}^{\text{NG}}=0.420$. Although, this partition reflects some information of the true partition, it fails to give the right number of communities.

Specifying $N_i=k_i$, $f(d)=e^{-(d/\sigma)^2}$ and estimating $d_{ij}$ by the Jaccard distance \cite{JaccardIndex, LevandowskyDistanceBetweenSets} between the $i$th and $j$th rows of $\mathbf{A}$, we apply Dist-Modularity optimization algorithm to this network and find two interesting partitions. One is shown in Fig.~\ref{fig3a}, which is obtained when $\sigma \in [0.61\bar{d}, 0.64\bar{d}]$ and $\bar{d}$ is the average distance between all node pairs. Note that only node 10 is classified wrongly. As can be seen form the figure, node 10 links to both of the two true groups with only one edge. The difference is that it directly links to the central node of the left group (node 34) while it links to the central node of the right group (node 1) with two hops. It is this minor difference that brings node 10 to the right community by our Dist-Modularity optimization algorithm. Nevertheless, this partition reflects most information of the club's fission. The other partition, obtained when $\sigma \in [2\bar{d}, +\infty]$, is shown in Fig.~\ref{fig3b}. Note that it is the same as that obtained by optimizing NG-Modularity. The reason is that, when the field range becomes long enough Dist-Modularity gradually reduces to NG-Modularity. It is worthwhile to mention that when $\sigma \in (0.61\bar{d}, 2\bar{d})$, the corresponding partitions are unstable, with one or several nodes's assignments deviating from the above two partitions.

\subsection{Bottlenose Dolphin Network}\label{sec4.3}
Now we turn to another famous network --- the bottlenose dolphin network that attracts considerable interests. This network was constructed by Lusseau from observations of 62 bottlenose dolphins living in Doubtful Sound over seven years \cite{LusseauDolphinNetwork1, LusseauDolphinNetwork2}. The nodes represent the dolphins and edges are placed between dolphin pairs that were observed together more often than expected by chance. During the observation period, one dolphin left the place temporarily and caused the fission of the dolphins into two groups of sizes 20 and 62. As shown in Fig.~\ref{fig4a}, the two groups are represented as square and circle symbols, respectively. Like Zachary's karate club network, the dolphin network is often used to test community detection algorithms, and the challenge is whether an algorithm can obtain the true partition of the two groups from the network structure.

Again, we specify $N_i=k_i$, $f(d)=e^{-(d/\sigma)^2}$ and estimate $d_{ij}$ by the Jaccard distance. Applying Dist-Modularity optimization algorithm to this network, we find two interesting partitions. The two-community partition shown in Fig.~\ref{fig4a} is obtained when $\sigma = 0.5\bar{d}$. This partition reflects most information of the dolphins' fission, as only node 40 is classified wrongly. The five-community partition shown in Fig.~\ref{fig4b} is obtained when $\sigma \in [0.9\bar{d}, +\infty]$. This partition is the same as that obtained by optimizing NG-Modularity, with the value of $\text{Q}^{\text{NG}}=0.529$ \cite{AgarwalModularityMaximizingMathematicalProgramming}.

In the above two networks, we use the same way to specify $N_i$, $f(d)$ and to estimate $d_{ij}$. Nevertheless, we have many other choices. For example, we have estimated $d_{ij}$ by the Euclidean distance, Minkowski distance \cite{DezaEncyclopediaDistances}, and the shortest distance \footnote{The shortest distance measures the length of the shortest path between two nodes.} Besides, we have tried $N_i=2m/n$ and $f(d)=1/\big(1+(d/\sigma)^2\big)$. Although these corresponding results are not shown here due to space constraints, they are quite similar to those described above. The reason is that, with the field theory as a basis for describing interactions between nodes, our null model captures the intrinsic feature of the observed network and serves as a competent reference.

\begin{figure}[!t]
\centering
\includegraphics[width=0.35\textwidth]{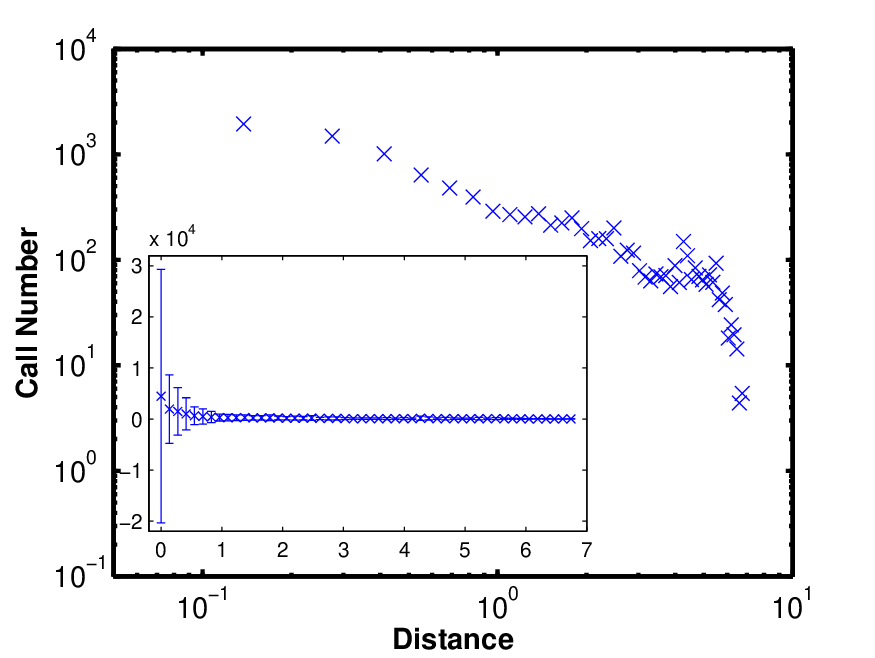}
\caption{\label{fig5} The number of calls between antennas decreases with their distance.}
\end{figure}

\subsection{Antenna-To-Antenna Network}\label{sec4.4}
Both the Zachary's karate club network and the bottlenose dolphin network are plain networks with only topology information. The third network in our experiments, namely the antenna-to-antenna network, is the one that has additional information on nodes. This network is based on anonymous records of mobile phone calls between five million of Orange's customers \footnote{Orange is the key brand of France Telecom, one of the world's leading telecommunications operators.} in Ivory Coast between Dec 1, 2011 and Apr 28, 2012 \cite{D4DData}. The nodes represent 1216 antennas which are associated with geographical position information, namely the coordinates of the two-dimensional space. The edges are placed between antennas that have communications, with the edge weight representing the numbers of calls. Note that this network is a temporal network \cite{HolmeTemporalNetworks}. It has ten consecutive slices and each slice represents a two-week period (all together accounts for the five-month period from Dec 1, 2011 to Apr 28, 2012).

We use $d_{ij}$ to denote the Euclidean distance between coordinates of antennas $v_i$ and $v_j$. we find that the number of calls between antennas decreases with their distance, as shown in Fig.~\ref{fig5}. In other words, the smaller of the distance between two nodes (i.e., the more similar of them), the higher of the probability that they are linked. Therefore, the similarity attraction feature applies to this network.

\begin{figure}[!t]
\centering
\includegraphics[width=0.50\textwidth, bb= 60 60 700 500]{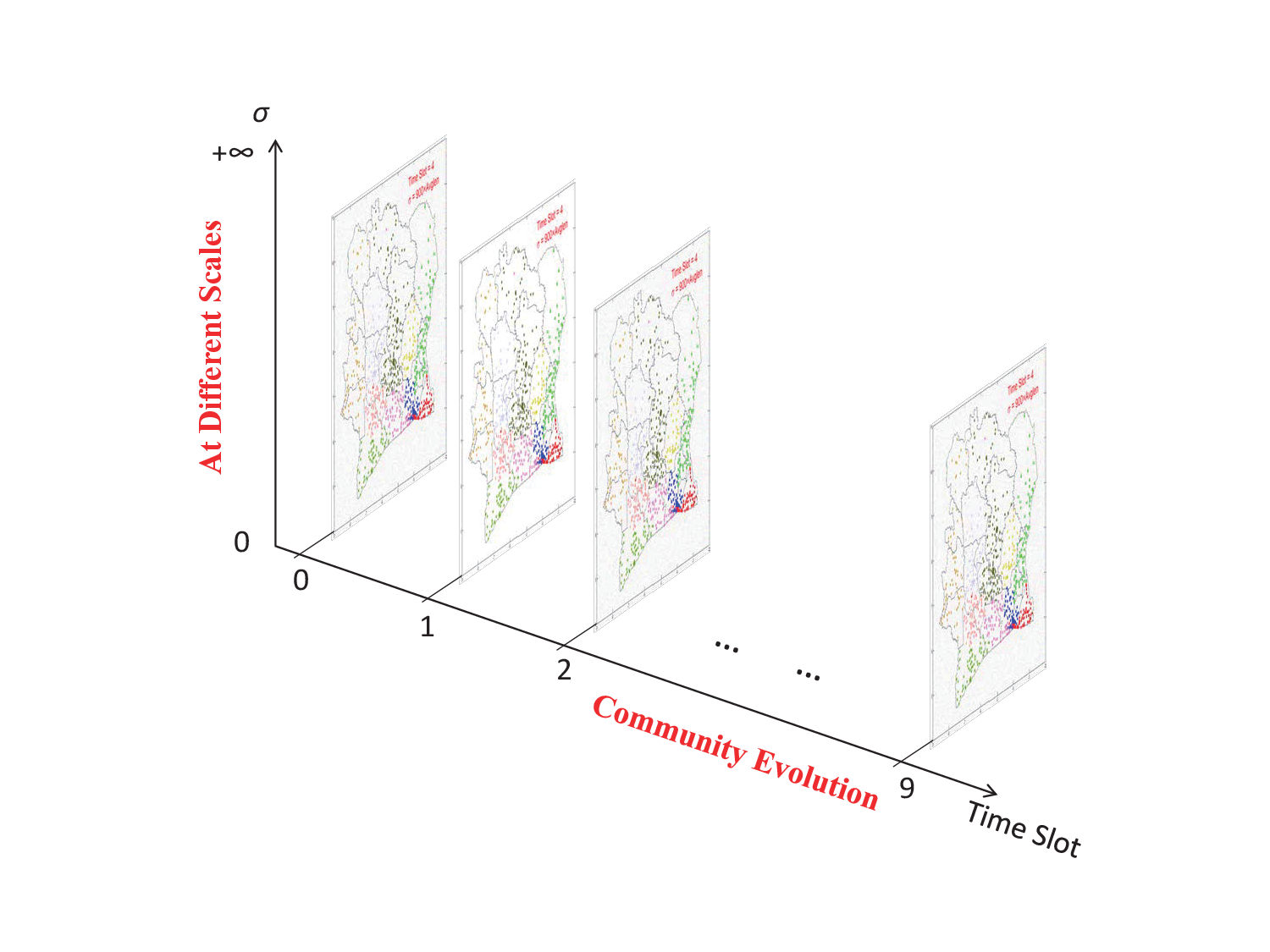}
\caption{\label{fig6} Explore communities in the antenna-to-antenna network.}
\end{figure}

Specifying $N_i=k_i$ and $f(d)=e^{-(d/\sigma)^2}$, we can explore the community structure by the Dist-Modularity optimization algorithm. As shown in Fig.~\ref{fig6}, we can explore communities at different scales along the $\sigma$ axis, and the community evolution along the time slot.

\begin{figure*}[!t]
\centering
\includegraphics[width=1\textwidth]{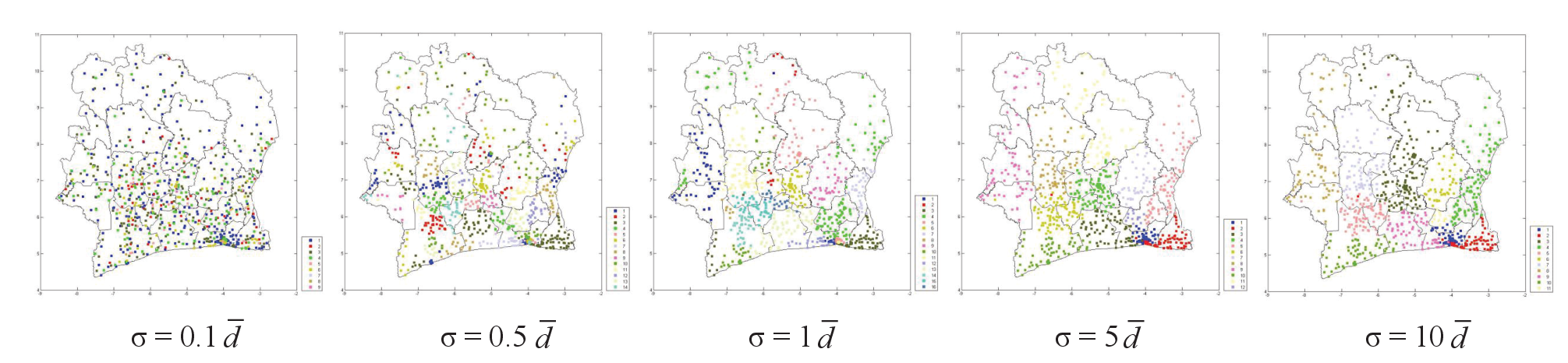}
\caption{\label{fig7} Explore communities at different scales as $\sigma$ increases from $0$ to $+\infty$.}
\end{figure*}

\begin{figure*}[!t]
\centering
\includegraphics[width=1\textwidth]{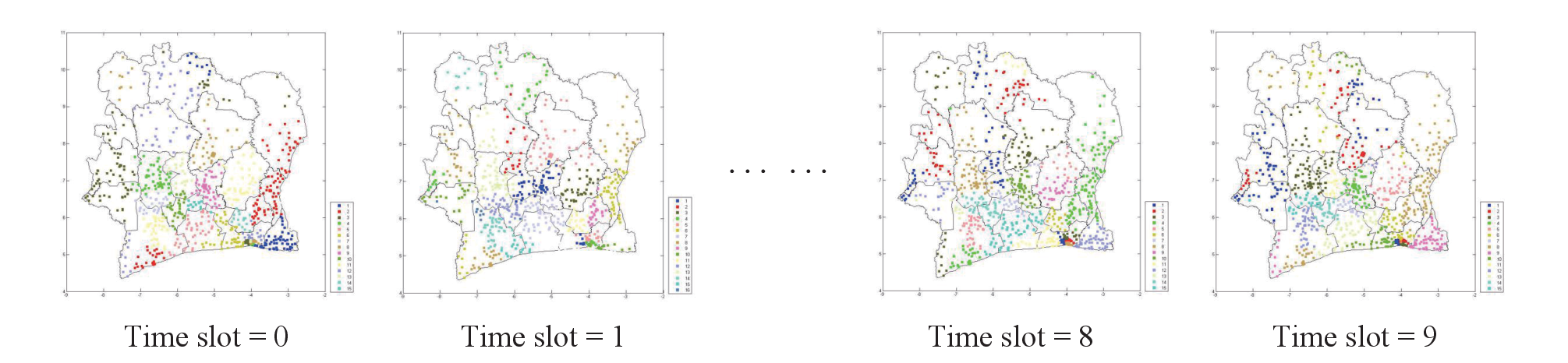}
\caption{\label{fig8} Explore the community evolution along the time slot.}
\end{figure*}

Fig.~\ref{fig7} shows the community structure in one of the network slice when $\sigma$ equals to $0.1\bar{d}$, $0.5\bar{d}$, $1\bar{d}$, $5\bar{d}$, and $10\bar{d}$, respectively. Fig.~\ref{fig8} shows the community evolution at $\sigma=1\bar{d}$. From Fig.~\ref{fig7} we can find that as $\sigma$ increases, the community structure gradually correlates with the geography. At $\sigma=1\bar{d}$, the border lines of the 19 regions of the country becomes a perfect classifier. Later, some of the communities are combined, resulting in coarser communities at $\sigma=10\bar{d}$. It can be expected that the community structure at $\sigma=10\bar{d}$ is quite similar to that obtained by optimizing NG-Modularity. Therefore, optimizing Dist-Modularity can find community structure at finer scales than by optimizing NG-Modularity. It is also interesting to notice that the scales of the community structure do not change with $\sigma$ linearly. The number of communities at $\sigma=0.1\bar{d}$ is 9. This number increases to 16 at $\sigma=1\bar{d}$ and finally falls to 11 at $\sigma=10\bar{d}$.

\section{Related Work}\label{sec5}
The study of community detection in networks has a long history. It is closely related to graph partitioning \cite{KernighanGraphPartitioning} in computer science, and hierarchical clustering \cite{ScottSocialNetworkAnalysis} in sociology. In the past decade, this study has attracted a great deal of interests and various methods were proposed \cite{FortunatoCommunityDetectionReview,NewmanCommunityReview,DanonCompareCommunityAlgo,LancichinettiCommunityAlgoAnalysis,LescovecCommunityAlgoComparison}. In particular, modularity optimization \cite{NewmanGreedy, ClausetFastGreedy, DuchExtremalOptimization, MedusSA, NewmanModularityEigenvectors, SchuetzMultistepGreedy, BlondelFastUnfolding, WakitaCommunityMegaScaleNetwork, LiuLPAmplus} is widely used despite its intrinsic limits \cite{FortunatoResolutionLimit, lancichinettiModularityLimit, GoodModularityDegeneracy}.

In many cases there exist multiple meaningful partitions of the network into communities, and these partitions correspond to different resolution scales. Within the modularity context, there are two main approaches for solving the multi-scale community detection problem. Arenas et al. control the resolution scale by adding a parameter that forms self-loops for each node \cite{ArenasDifferentResolutionLevels}. Reichardt and Bornholdt control the resolution scale by adding a parameter in front of the null model term in the definition of modularity \cite{ReichardtStatisticalMechanicsCommunityDetection}. However, it has been recently demonstrated that these two methods are intrinsically deficient and their use do not produce reliable solutions \cite{lancichinettiModularityLimit}.

Recently, some researchers considered community detection in networks with additional information on nodes. Expert et al. \cite{ExpertSpatialCommunity} and Cerina et al. \cite{CerinaSpatialCorrelation}, respectively, proposed a method by factoring out the effect of space in spatial networks where the geographical information on nodes is available \cite{BarthelemySpatialNetworkReview}. Yang et al. devised a discriminative approach for combining the edge and node attributes \cite{YangCommunityDetectionCombiningLinkContent}. Sachan et al. proposed generative models that incorporate contents and interactions for discovering communities in social networks \cite{SachanContentInteractionsDiscoveringCommunities}.

\section{Conclusion}\label{sec6}
With the field theory as a basis for describing interactions between nodes, we propose a null model that captures the similarity attraction feature of real-world networks. Taking this null model as a reference for comparing the fraction of within-community edges with the observed network, we create a framework for generating a family of Dist-Modularity adapted to various networks, including networks with addition information on nodes. Dist-Modularity, which incorporates NG-Modularity as a special case, can be used to detect communities at different scales.

One of the notable drawback of NG-Modularity is the resolution limits, which leads to the incapability of identifying small communities in large-scale networks. The resolution limit is attributed to the globalization of the null model, which assumes that any node can link to any other node of the network. Our null model, with the similarity attraction feature built in, in some sense restricts but doest not completely prohibit such globalization. We show that optimizing Dist-Modularity finds community structure at finer scales than by optimizing NG-Modularity. However, this problem still needs further investigation and this is left for our future work.

\section{Acknowledgments}
This work was partly funded by CREST, JST and by NSFC under grant number 61203154.

\bibliographystyle{abbrv}
\bibliography{myRef15}

\end{document}